\documentclass[aps,prl,twocolumn,showpacs,amsmath,amssymb]{revtex4}
\usepackage{epsf}
\usepackage{graphicx}
\usepackage{bm}

\begin{document}

\title{Tunneling without tunneling: wavefunction reduction in a
mesoscopic qubit}

\author{James A. Nesteroff and Dmitri V. Averin}

\affiliation{Department of Physics and Astronomy, Stony Brook
University, SUNY, Stony Brook, NY 11794-3800 }

\date{\today}

\begin{abstract}

The transformation cycle and associated inequality are suggested for
the basic demonstration of the wavefunction reduction in a
mesoscopic qubit in measurements with quantum-limited detectors.
Violation of the inequality would show directly that the qubit state
changes in a way dictated by the probabilistic nature of the
wavefunction and inconsistent with the dynamics of the
Schr\"{o}dinger equation: the qubit tunnels through an infinitely
large barrier. Estimates show that the transformation cycle is
within the reach of current experiments with superconducting qubits.

\end{abstract}


\maketitle

Can a quantum particle tunnel through a barrier which has vanishing
transparency? Immediate answer to this question is ``no'' as follows
from the elementary properties of the Schr\"{o}dinger equation. More
careful consideration should of course remind that evolution
according to the Schr\"{o}dinger equation is not the only way for a
state of a quantum particle to change in time. Probabilistic nature
of the wavefunction implies that it can also evolve due to the
``wavefunction reduction'': random process of realization of one
specific outcome of a measurement. In this process, the wavefunction
of the system changes coherently for any given outcome of the
measurement -- see, e.g, \cite{b1,b2,b3}, if measured with the
``morally best'' \cite{b2,b3} or, in a more modern and descriptive
language, quantum-limited detector. Such an evolution of the
measured system can contradict the Schr\"{o}dinger equation despite
the fact that the dynamics of the measurement process as a whole is
governed by this equation. It can be described formally as generic
quantum operation within the approach based on positive
operator-valued measures (POVM) \cite{b4}. All ``counter-intuitive''
quantum-mechanical phenomena arise from the wavefunction reduction.
The best known example is given by the EPR correlations which
violate the principle of ``no action at-a-distance'' as quantified
by the Bell's inequalities. They appear for specific random outcomes
of the local spin measurements. On average, there is no
action-at-a-distance, since the correlations do not violate the
relativistic causality. In mesoscopic solid-state qubits, proposed
and/or observed manifestations of the wavefunction reduction include
violations of the temporal Bell inequalities
\cite{b10,b11,b12,b12*}, measurements of the ``weak values'' of the
operators \cite{r4,r5}, stochastic reversal of the wavefunction
reduction \cite{r1}. The purpose of this work is to suggest a
sequence of quantum transformations and the corresponding inequality
which illustrates directly the contradiction between the
wavefunction reduction and the Schr\"{o}dinger dynamics by violating
the very basic intuition that a particle can not tunnel through an
infinitely large barrier. For mesoscopic structures with their small
geometric dimensions, violation of this intuition would provide,
arguably, more dramatic illustration of the wavefunction reduction
than the non-locality of conventional Bell's inequalities.

The system we consider is a qubit with the two basis states
$|j\rangle$, $j = 0,1$, distinguished by the average values of some
quantity $x$, for instance, electric charge or magnetic flux
\cite{b22,b23,b24,b25} in the case of superconducting qubits. As the
simplest example, $x$ can be viewed as a position of an individual
particle (electron in coupled quantum dots \cite{b19,b20}, Cooper
pair on a superconducting island \cite{b22,b23}, FQHE quasiparticle
in a system of two quantum antidots \cite{b17,b30}, or ultracold
atom in a BEC junction \cite{b31,b32}) which can be localized on the
opposite sides of a tunnel barrier separating the states $|j\rangle$
and creating tunnel amplitude $\Delta > 0$ between them (Fig.~1).
The coordinate $x$ is measured by a detector which converts the
information about $x$ into the classical output $q$. The detector is
characterized by the probabilities $w_j(q)$ of producing the output
$q$, when the qubit is in the state $|j\rangle$. For instance, in
the example of the quantum-point-contact (QPC) detector (see, e.g.,
\cite{b9} and references therein), the qubit controls the scattering
characteristics of electrons in the contact, and therefore the
current $I$ flowing through it (Fig.~1). In this case, the output
$q$ is the total charge transferred through the contact during the
time $\tau$ of the measurement.

\begin{figure}[htc]
\setlength{\unitlength}{1.0in}
\begin{picture}(4.2,1.3)
\put(0.0,-.1){\epsfxsize3.3in\epsfbox{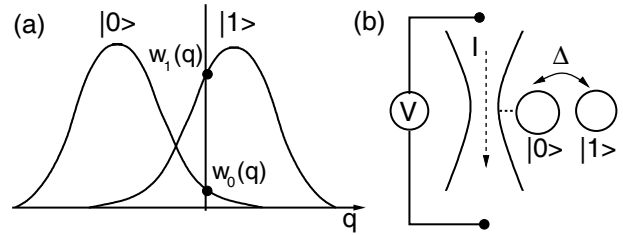}}
\end{picture}
\caption{Illustration of the qubit measurement process. (a) Typical
probability distributions $w_j(q)$ of a detector output $q$ for the
qubit in the state $|j\rangle$,  $j=0,1$. (b) Schematics of the
particular measurement setup based on the QPC detector measuring
mesoscopic ``charge'' qubit. In this case, the qubit states
$|j\rangle$ differ by position of some elementary charge which
controls the QPC transmission properties, and therefore, the current
$I$ in the contact driven by the applied voltage $V$. }
\end{figure}

The goal of the transformation cycle developed in this work is to
demonstrate that the evolution of the wavefunction in the
measurement process is ``real'' to the same extent as the dynamics
of wavefunction governed by the Schr\"{o}dinger equation, i.e., it
describes evolution of the physical quantities and not only
information about them. This is accomplished by combining the two
types of evolution in one cycle arranged so that they completely
compensate each other, leaving the initial qubit state unchanged.
The first part of the cycle is the wavefunction reduction in a weak
quantum-limited measurement which mimics the tunneling between the
qubit states $|j\rangle$ through infinitely large barrier with
$\Delta=0$. The second part is the regular tunneling with $\Delta
\neq 0$. The fact that no charge or flux is transferred through the
tunneling barrier in the whole cycle means that the wavefunction
reduction induces tunneling even without the corresponding tunneling
amplitude.

The starting point of the cycle is a qubit with vanishing average
bias $\epsilon=0$ between the states $|j\rangle$. The Hamiltonian of
such a qubit is
\begin{equation}
H = -( v\sigma_z +\Delta\sigma_x)/2 \, , \label{e1}
\end{equation}
where $\sigma_{x,z}$ are the Pauli matrices, and the bias $v$
represents the low-frequency noise characteristic for mesoscopic
solid-state qubits \cite{b33,b34,b35,b36}. Due to assumed weak, but
unavoidable, relaxation, the qubit at low-temperature $T\ll \Delta$
is in the instantaneous ground state of (\ref{e1}) with the density
matrix $\rho_i$ in the $\sigma_z$ basis
\begin{equation}
(I) \;\;\;\;\;\; \rho_i = \left( \!\!\!  \begin{array}{cc} c_0^2 &
c_0c_1 \\ c_0c_1 & c_1^2 \end{array} \!\!\!  \right) , \label{e2}
\end{equation}
where the probability amplitudes $c_j$ are $c_{0,1} = [(1 \pm
v/\Omega)/2]^{1/2}$, and $\Omega = (v^2 + \Delta^2)^{1/2}$. For
$\Delta \gg v_0$, where $v_0$ is the r.m.s. magnitude of noise $v$,
the state (\ref{e2}) is the ideal version of initial state for our
transformation cycle, the eigenstate $\sigma_x=1$ of the $\sigma_x$
operator with $c_{0,1} =1/\sqrt{2}$. However, optimization of the
cycle as a whole discussed below can require keeping the ratio
$\Delta/v_0$ finite, which reduces the noise-induced dephasing.

The cycle begins by rapidly (on the scale $\hbar/v_0$) raising the
tunnel barrier so that the tunnel amplitude vanishes, $\Delta
\rightarrow 0$, while the qubit wavefunction remains distributed
between the states $|j\rangle$. The next step is a weak measurement
of the $\sigma_z$ operator performed on the qubit state (\ref{e2})
with a quantum-limited detector. Qualitatively, this means that the
fluctuations underlying the probability distributions $w_j(q)$ of
the detector output are themselves quantum, so that the evolution of
the detector+qubit system leading to any given value $q$ of the
output is quantum coherent. An example of the quantum-limited
detector is the QPC \cite{b7,b9} mentioned above (Fig.~1b), in which
the distribution of the transferred charge $q$ is created by the
quantum-coherent scattering of electrons at the contact. The
information about the qubit state gained by the measurement depends
on the observed output $q$ (Fig.~1a), and implies that the qubit
wavefunction amplitudes $c_j$ evolve into the $q$-dependent values
$c_j(q)$ \cite{b7,b8}:
\begin{equation}
c_j(q) = \frac{c_j\sqrt{w_j(q)}}{\sqrt{w_0(q)|c_0|^2 +
w_1(q)|c_1|^2}}. \label{e3}
\end{equation}

If the detector is not strictly quantum-limited, it induces partial
dephasing of the qubit state even for a given specific detector
output $q$. Physically, this dephasing is caused by the loss of
information in the measurement process and can be described in
general by the suppression factor $e^{-\eta}$ of the non-diagonal
elements of the qubit density matrix $\rho$ in the measurement
basis. The qubit density matrix after weak $\sigma_z$ measurement
the is
\begin{equation}
(II) \;\;\;\;\;\; \rho_m = \left( \!\!\! \begin{array}{c}
c_0^2 (q)\, , \;\;\;\; c_0(q)c_1(q)e^{-\eta+ i \varphi(\tau)} \\
c_0(q)c_1(q) e^{-\eta- i \varphi(\tau) }, \;\;\;\; c_1^2(q)
\end{array} \!\!\! \right) . \label{e4} \end{equation}
The factor $e^{i \varphi(\tau)}$ here represents the noise-induced
phase $\varphi(\tau)= \int_0^{\tau} dt v(t)/\hbar$ during the
measurement time $\tau$.

The aim of the next step of the cycle is to reverse the changes in
the qubit amplitudes $c_j$ due to redistribution through the tunnel
barrier with vanishing transparency in the reduction process
(\ref{e3}) by regular tunneling. This is done by creating a
non-vanishing tunneling amplitude for some appropriate period of
time, i.e., realizing a fraction of the regular coherent
oscillations in which the charge or flux goes back and forth between
the qubit basis states. In the situation with no disturbances
(vanishing noise and quantum-limited detector), this can be done
precisely, returning the qubit from the state
\[|\psi_m \rangle= [\sqrt{w_0(q)}|0\rangle+\sqrt{w_1(q)}
|1\rangle]/\sqrt{w_0(q)+w_1(q)} \, .\]
obtained as a result of the measurement, to the initial state
$|\psi_i\rangle=[|0\rangle+ |1\rangle]/\sqrt{2}$ before the
measurement. Writing the amplitudes of $|\psi_m \rangle$ as $\cos
(\theta/2)$ and $\sin (\theta/2)$, one can see that the required
transformation is the rotation about the $y$ axis. For mesoscopic
qubits, such a rotation corresponds to complex tunnel amplitude
$\Delta'$ with $\arg \Delta' = \pi/2$, and the rotation angle is
\begin{equation}
\int dt \frac{|\Delta'(t)|}{\hbar}= \frac{\pi}{2} - \theta \, ,
\;\;\;\; \theta = 2 \tan^{-1}\left(\frac{w_1(q)}{w_0(q)}
\right)^{1/2} . \label{e6}
\end{equation}

Typically, the qubit structure allows only for the real tunnel
amplitude $\Delta$ -- see the Hamiltonian (\ref{e1}), which realizes
the $x$-axis rotations $R_x= \exp \{i \sigma_x \int \Delta(t)dt/2
\hbar \}$ of the qubit. In this case, the $y$-axis rotation
(\ref{e6}), $R_y= \exp \{-i \sigma_y \int |\Delta'(t)|dt/2 \hbar
\}$, can be simulated directly by the $x$-rotation of the same
magnitude, if it is preceded and followed by the $z$-axis rotations:
$R_y=R_z^{-1} R_x R_z$. The $z$-rotations $R_z^{\pm 1} =\exp \{\pm i
\sigma_z \pi/4 \}$ are created by the pulses of the qubit bias:
$\int \epsilon (t)dt/\hbar = \pm \pi/2 $. Such a three-step sequence
is simplified by interchanging the order and magnitude of rotations:
first the $x$-, then one $z$-rotation:
\begin{equation}
(III) \;\;\;\;\; \int dt \frac{\Delta(t)}{\hbar}= \frac{\pi}{2} \, ,
\;\;\; \int dt \frac{\epsilon (t)}{\hbar} = \frac{\pi}{2}- \theta \,
. \label{e7} \end{equation}
Under the experimentally-realistic assumption that the measurement
takes much longer time than control pulses, the noise-induced
distortions of the qubit states is dominated by the phase
accumulation in Eq.~(\ref{e4}). The qubit density matrix after the
pulses (\ref{e7}) then is
\begin{equation}
\rho =U \rho_m U^{\dagger} \, , \;\;\;\; U=R_z(\pi/2-
\theta)R_x(\pi/2) \, .\label{e8}
\end{equation}

As the last step, one needs to check whether the qubit is brought
back to the initial state $\sigma_x = 1$. This is done directly by
strong measurement of $\sigma_x$, which gives, upon sufficiently
large number of cycles, the error probability $p$ of the qubit
reaching the wrong state $\sigma_x = -1$:
\begin{equation}
(IV) \;\;\;\;\; p= \mbox{Tr} [P\rho] \, , \;\;\;\; P= \frac{1}{2}
\left( \!\!\!  \begin{array}{cc} 1 & -1 \\ -1 & 1 \end{array} \!\!\!
\right) . \label{e9}
\end{equation}
If $p=0$, the cycle $(I)-(IV)$ returns the qubit with certainty to
the state $\sigma_x = 1$. This means that for any finite measurement
strength, when the distributions $w_0(q)$ and  $w_1(q)$ are not
identical, the transfer of the qubit amplitudes due to the
wavefunction reduction with $\Delta=0$ is precisely compensated for
by a fraction of a period of coherent qubit oscillations with
$\Delta \neq 0$. Since the oscillations actually transfer the charge
or flux between the two qubit basis states, the fact that the cycle
is closed shows that the qubit evolution in the wavefunction
reduction can not be interpreted only as the changes in our
knowledge of the qubit state, but rather involve actual transfer of
charge or flux without the tunneling amplitude.

In the presence of the finite noise and detector non-ideality, the
error probability $p$ is non-vanishing. Still, the probability $1-p$
of the closed cycle can be larger than the value explainable by the
classical description of the measurement process, demonstrating that
the actual qubit evolution in measurement is governed by the
wavefunction reduction (\ref{e3}). The alternative classical
description would be based on the assumption that the process of
switching off the tunneling amplitude $\Delta$ does not leave the
qubit in the state (\ref{e2}) but localizes it in one of the basis
states on one or the other side of the tunnel barrier with some
undetermined probability $r$, so that the qubit density matrix is $
\rho^{(cl)}= r|0\rangle \langle 0| + (1-r)|1\rangle \langle 1|$. The
qubit state is then ``objectively'' well-defined and coincides with
one of the basis states. It is, however, unknown, and the
measurement provides information about this unknown state changing
the probability $r$:
\[ r \rightarrow r(q)=rw_0(q)/(rw_0(q) + (1-r)w_1(q)) \, .\]
Applying the same transformations as to the state (\ref{e2}) to $
\rho^{(cl)} (q)= r(q)|0\rangle \langle 0| + (1-r(q))|1\rangle
\langle 1|$, one finds the error probability $p^{(cl)}(q)$ given the
measurement outcome $q$. Averaging over the probability $\sigma(q)=
r w_0(q) + (1-r) w_2(q)$ of different $q$'s, one obtains the total
classical error probability
\begin{equation}
p^{(cl)} = \int dq \, \frac{w_0(q)w_1(q) }{w_0(q) + w_1(q)} \, .
\label{c13} \end{equation}
Equation (\ref{c13}) shows that the probability $p^{(cl)}$ is
independent of the assumed initial probability $r$, i.e. of noise
$v$, and of the degree of detector non-ideality. The probability
(\ref{c13}) characterizes the measurement strength, with
$p^{(cl)}\rightarrow 0$ for strong projective measurements, when
$w_0(q)w_1(q)=0$, i.e. the measurement provides definite information
about the qubit state. For weak measurements, the two distributions
$w_{0,1}(q)$ nearly coincide and $p^{(cl)}\rightarrow 1/2$.

To see the incompatibility of the classical description with the
actual qubit evolution, one needs to compare $p^{(cl)}$ (\ref{c13})
with the probability $p$ found from Eqs.~(\ref{e8}) and (\ref{e9}),
after averaging over the detector output:
\begin{equation}
p = p^{(cl)} \left[1-e^{-\eta} \, F \right] . \label{e12}
\end{equation}
The factor $F$ here describes effect of the noise $v$:
\begin{equation}
F\equiv \Big\langle \frac{\Delta}{ (v^2 + \Delta^2)^{1/2}} \cos \!\!
\int_0^{\tau} \! ( v(t)dt/\hbar ) \Big\rangle \, . \label{e14}
\end{equation}
Equation (\ref{e12}) shows that if the noise and the detector
non-ideality are not very large: $F\rightarrow 1$ and $\eta
\rightarrow 0$, the errors of the quantum cycle are suppressed
regardless of the measurement strength characterized by $p^{(cl)}$,
and it should be possible to observe that
\begin{equation} p< p^{(cl)}. \label{e13} \end{equation}
Demonstration of this inequality would mean that a larger number of
the transformation cycles are closed than can be explained
classically, without invoking the transfer of the amplitude $c_j$
across the infinite barrier. Note that Eq.~(\ref{e12}) is obtained
for essentially arbitrary noise model, and is therefore valid in the
general case of the quantum noise that corresponds to entanglement
of the qubit (\ref{e1}) with an arbitrary quantum system averaged
out in Eq.~(\ref{e14}).

To make the limitations on the noise more quantitative, we adopt the
usual model of the low-frequency noise (see, e.g., \cite{b34,b36})
as a classical Gaussian random variable static on the time scale of
the cycle, which gives
\begin{equation}
F= \frac{1}{\sqrt{2\pi} v_0} \int dv e^{-\frac{v^2}{2v_0^2}} \,
\frac{\Delta \cos (v\tau/\hbar) }{(v^2 + \Delta^2)^{1/2}} \, .
\label{e17} \end{equation}
Equation (\ref{e17}) can be evaluated analytically in several
limits. For $\Delta \gg v_0$, the noise effect on the initial state
disappears, and Eq.~(\ref{e17}) reduces for arbitrary
$v_0\tau/\hbar$ to pure dephasing
\begin{equation}
F= e^{-v_0^2\tau^2 /2\hbar^2} . \label{e18} \end{equation}
In the opposite limit of negligible dephasing, $v_0\tau/\hbar \ll
1$, or if dephasing is eliminated by a spin echo technique,
\begin{eqnarray}
F = \sqrt{2a/\pi} \, e^{a} K_0 (a)\, , \;\;\; a \equiv \Delta^2/(4
v_0^2)\, , \label{e15}
\end{eqnarray}
where $K_0$ is the modified Bessel function. This equation describes
the transition from $F\simeq 1$ for $\Delta \gg v_0$, to $F\simeq
(\Delta/v_0)\, \ln (v_0/\Delta)$ for $\Delta \ll v_0 $. For $\Delta
\ll v_0$ and $\tau \neq 0$, one can neglect the exponential factor
in (\ref{e17}) to get:
\begin{equation}
F= (2/\pi)^{1/2} (\Delta/v_0) K_0 ( \Delta \tau/\hbar) \, .
\label{e19} \end{equation}

\begin{figure}[h]
\includegraphics[width=0.36\textwidth]{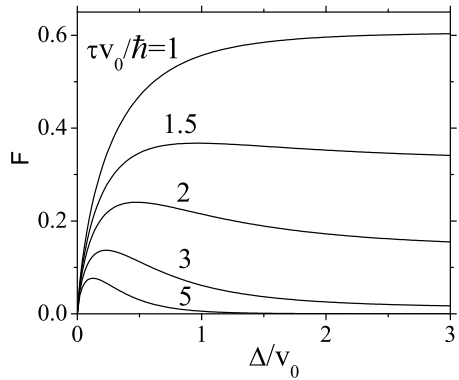}
\vspace{-.1in}
\caption{The noise suppression factor \protect (\ref{e17}) as a
function of the $\Delta/v_0$ ratio for several values of the noise
strength or quality parameter $v_0 \tau/\hbar$ of the qubit-detector
system. For the discussion, see main text. \label{f4}}
\end{figure}

The strength of the qubit coupling both to the detector and the
noise depends typically on the same qubit parameters. This means
that optimization for maximum $F$ should be done for fixed values of
the $v_0 \tau$ product which represents a more fundamental quality
of the qubit-detector system. Figure \ref{f4} shows $F$ as a
function of the $\Delta/v_0$ ratio for several values of $v_0
\tau/\hbar$ calculated numerically from Eq.~(\ref{e17}). In the
large-$\Delta$ limit, the curves saturate at $F$ given by
Eq.~(\ref{e18}), which is exponentially small for strong noise. For
$v_0 \tau/\hbar \geq 2$, the $\Delta$-dependence of $F$ is
non-monotonic, with the maximum at small $\Delta/v_0$ representing
suppression of dephasing by small $\Delta$. Although the reduction
of $\Delta$ increases the noise effect on the initial state, it
effectively suppresses dephasing: the instances of stronger noise
which would result in a stronger dephasing become irrelevant, since
the qubit state is already localized for them in one of the states
$|j\rangle$ and does not contribute to the wavefunction reduction.
On the other hand, the probability of obtaining realizations with
small noise scales as $\Delta/v_0$, and $F$ is also suppressed for
$\Delta\rightarrow 0$. This gives the peak of $F$ at small
$\Delta/v_0$ seen in Fig.~\ref{f4} and  for $v_0 \tau/\hbar\gg 1$,
described analytically by Eq.~(\ref{e19}). This peak may be useful
in experimental realization of the cycle discussed in this work.

In general, realization of this cycle and observation of the
inequality (\ref{e13}) is not very far from the possibilities of the
current experiments with superconducting qubits. Indeed, Figure
\ref{f4} shows that a noticeable difference between the classical
and quantum error probabilities can be obtained even for the cycle
times $\tau$ on the order of $\sim 3 $ dephasing times (\ref{e18})
$\hbar/v_0$. Experimental dephasing times for the qubit states that
differ by the average flux or charge values are about 5 ns -- see,
e.g., \cite{b25,b15}, giving roughly 15 ns interval for performing
three operations of the cycle when the qubit is subject to
dephasing: measurement and two compensating pulses (\ref{e7}).
Simple qubit control pulses are regularly performed on the
few-nanosecond time scale, and can be fit into this interval. The
new requirement presented by the cycle described above is the need
to perform a variant of the feed-back control, when the applied
pulses depend on the result of the previous qubit measurement.
Necessary rapid non-destructive read-out could be developed, e.g.,
using the superconductor qubit control circuits \cite{b8}.

In summary, we have proposed a transformation cycle aimed at the
very basic demonstration of the two main features of the
wavefunction reduction in quantum measurements using mesoscopic
solid-state qubits. The wavefunction can evolve in a way that
explicitly contradicts the dynamics of the Sch\"{o}dinger equation:
a particle can be transferred through an infinitely large barrier.
On the other hand, to the same extent as the Sch\"{o}dinger
equation, the reduction affects not only the probability
distributions of a dynamic variable (i.e., electric charge or
magnetic flux), but the variable itself. Although the particle
transfer through the infinite barrier in this process is not
identical to the regular Sch\"{o}dinger-equation tunneling, e.g.,
there is no a corresponding current operator, the cycle developed in
this work demonstrates that such a transfer can still take place and
should be observable in current experiments with superconducting
qubits.

D.V.A. would like to thank L.Y. Gorelik, A.N. Korotkov, and V.K.
Semenov for useful discussions. This work was supported in part by
the NSA under ARO contract W911NF-06-1-217.

\end{document}